\documentclass[12pt]{article}
\title{CHIRAL SYMMETRY IN LINEAR SIGMA MODEL IN MAGNETIC ENVIRONMENT}
\author{Ashok Goyal\thanks{E--mail : agoyal@ducos.ernet.in}, 
and
Meenu Dahiya\thanks{E--mail : meenu@ducos.ernet.in}  \\
	{\em Department of Physics and Astrophysics,} \\
	{\em University of Delhi, Delhi-110 007, India.} \\
        {\em InterUniversity Centre for Astronomy and Astrophysics,} \\
        {\em Ganeshkhind, Pune 411007 , India.} \\
        }

\begin{document}
\maketitle
\large
\begin{abstract}
We study the chiral symmetry structure in a linear sigma model with 
fermions in the presence of an external, uniform magnetic field in
the 'effective potential' approach at the one loop level.
We also study the chiral phase transition as a function of density in
the  core of magnetized neutron stars.
\end{abstract}
\pagebreak
In the absence of quark mass matrix, QCD is invariant under chiral
transformation at the lagrangian level.
However, the dynamics of QCD are expected to be such that chiral
symmetry is spontaneously broken with the vacuum state having a
nonzero quark-antiquark condensate $ \left\langle 0\left\vert\bar
qq\right\vert 0 \right\rangle $  and the Goldstone theorem then
requires the existence  of approximately massless pseudo-scalar mesons
in the hadron spectrum. 
At high temperatures and/or at high densities,the quark condensates
are expected to melt at some critical point and chiral symmetry is 
restored \cite{brown}.
Chiral phase transition and phenomenological consequences in the form
of experimentally observable signatures have been extensively
discussed in the literature \cite{harris}.
It has also been suggested \cite{linde1,salam1} that systems with
spontaneously broken symmetries may make a transition from broken
symmetric to restored symmetric phase in the presence of external
fields.
It has been shown \cite{salam2,linde2} that there exist some realistic
models where the symmetry restoration takes place.
In QED uniform, external static magnetic field is known to break
chiral symmetry dynamically at weak gauge couplings \cite{kikuchi}.

Large magnetic fields with  strengths upto $10^{18}$ gauss 
 have been conceived to exist \cite{ternov} at the time of supernova collapse
inside neutron stars and in other astrophysical compact objects  and
in the early Universe. 
Effect of such a strong magnetic field on 
chiral phase transition is  thus of great interest for baryon free
quark matter in the early universe and for high density baryon matter
in the core of neutron stars. 
  To study chiral phase transition in QCD we need a nonperturbative
treatment.
 Lattice techniques and the Schwinger-Dyson equations provide
specially powerful methods to study the chiral structure of QCD
\cite{miransky}.
Klevansky and Lemmer \cite{klevansky} studied the chiral symmetry behaviour of hadronic 
system described by Nambu-Jona-Lasinio model minimally coupled to a constant
electro-magnetic field. Solving the gap equation they found that whereas a 
constant electric field restores chiral symmetry, a constant magnetic field
inhibits the phase transition by stabilizing the chirally broken vacuum state.
This conclusion was confirmed by later studies of the NJL model \cite{klimenko}.
Shushpanov and Smilga \cite{smilga} considered QCD with the massless flavors in the 
leading order of chiral perturbation theory and studied the dependence of
quark condensate on an external magnetic field by studying the schwinger-
Dyson equation and showed that an external magnetic field increases the
condensate.   

A particularly attractive frame work to study such systems is the 
linear sigma model originally proposed as a model for strong nuclear 
interactions \cite{gellmann}. We will consider this as an effective model
for low energy phase of QCD and will examine the chiral symmetry properties 
at finite density and in the presence of external magnetic field. To fix 
ideas we consider a two flavor $SU(2) \times 
SU(2)$
chiral quark model given by the lagrangian.
\begin{equation}
\mathcal{L} =  i \bar{\psi} \gamma^\mu \partial_\mu \psi
- g \bar{\psi}( \sigma+ i \gamma_5  \vec{\tau} . \vec{\pi}) \psi 
+ {1 \over 2} (\partial_\mu \sigma)^2 
+ {1 \over 2} (\partial_\mu \vec{\pi})^2 - U(\sigma,\vec{\pi})
\end{equation} 
where $\psi$  is the quark field $\sigma$ and $\pi$ are the set of four scalar
fields and g is the quark meson coupling constant. The potential
 U$(\sigma ,\vec{\pi})$ is
given by \  
\begin{equation}
U( {\sigma}, \vec{\pi})= - \frac{1}{2} \mu^2 ( \sigma^2+ \vec{\pi}^2)+\frac{1}
{4}{\lambda} ( {\sigma}^2+ \vec{\pi}^2)^2
\end{equation}
For ${\mu}^2>0$ chiral symmetry is spontaneously broken. The $\sigma$ field
can be used to represent the quark condensate, the order parameter for chiral 
phase transition and the pions are  the Goldstone bosons. At the
tree level the sigma , pion and the quark masses are given by
\begin{equation}
m_{\sigma}^2=3{\lambda} {\sigma}_{cl}^2-{\mu}^2    ;    m_{\pi}^2={\lambda} {\sigma}_{cl}^2-{\mu}^2   ;    m_{\psi}^2 = g {\sigma}_{cl}
\end{equation}
 where ${\sigma}_{cl}^2=\frac{\mu^2}{\lambda} = f_{\pi}^2 $.
 An elegant and efficient way to study symmetry properties of the vacuum at 
finite temperature, density and in the presence of external fields is through
the ``Effective Potential'' approach discussed extensively in the literature
\cite{kirshnitz}. We will compute here, in the one loop approximation, the
effective potential in the presence of external magnetic field which is 
defined through an effective action $\Gamma(\sigma,B)$
 which is the generating functional of the one particle irreducible
 graphs. The effective potential is then given by 
\begin{equation}
V_{eff}(\sigma,B)=V_{0}(\sigma)+V_{1}(\sigma,B)
\end{equation}
where $V_{1}(\sigma,B)$ is obtained from the propagator function $G(\sigma,
B)$ by the usual relation
 $V_{1}(\sigma,B)=-\frac{1}{2 i} Tr \log G(\sigma,B)$.

 Alternatively one can compute the shift
in the vacuum energy density due to zero-point oscillations of the fields 
considered as an ensemble of harmonic oscillators \cite{kirshnitz}. We thus
require energy eigenvalues(excitations) of particles in the magnetic field,
which can be easily obtained, and in the absence of anomalous magnetic moment
for uniform static magnetic field in the z-direction for a particle of mass
M,charge q and spin J, are given by \cite{wu-yang} 
\begin{equation}
\label{one}
E(k_{z},n,J_{z})={(k_{z}^2+M^2+(2 n+1-sign(q)\,\,j_{z}) \left\vert 
q \right\vert B)}^2
\end{equation}
where n represents the landau level.
Contribution of scalar particles of mass M to $V_1(M^2)$ after wick rotation is
thus given by 
\begin{equation}
V_{1}(M^2)=\frac{1}{2}\int \frac{d^4 k_{e}}{(2 \pi)^4} ln (k_{e}^2+M^2-i \epsilon)
\end{equation}
In the presence of magnetic field, all we need to do is to replace the phase
space integral
 $\int \frac{d^4 k_{e}}{(2 \pi)^4} $  by  $\frac{e B}{2 \pi}
 \sum_{n=0}^\infty \frac{d^2 k_{e}}{(2 \pi)^2}$   
and the energy by expression (\ref{one})  for charged particles.
For a scalar field of charge $\pm e $, we thus have
\begin{eqnarray}
\label{three}
V_{1}(M^2,B)&=& \frac{ e B}{4 \pi}
\sum_{n=0}^\infty \int \frac{d^2 k_e}{(2 \pi)^2} \ ln (k_e^2+(2 n+1) 
e B+ M^2)
\nonumber \\
&=&-\frac{ e B}{4 \pi}\,\,  \frac{\partial}{\partial\alpha}
\,\,   \frac{\Gamma(\alpha-d/2)}{\Gamma(\alpha) (4\pi)^{d/2}} 
\nonumber \\
&&\sum_{n=0}^\infty \frac{1}{(M^2+e B+2 neB)^{\alpha-d/2}} \Bigg\vert_
{\alpha=0,d=2}
\nonumber \\
&=&-\frac{e B}{4 \pi}\,\, \lim_{\alpha \rightarrow 0}\,\, \frac
{\partial}{\partial\alpha}\,\, \frac{\Gamma(\alpha-\frac{d}{2})}{\Gamma
(\alpha) (4 \pi)^{\alpha-\frac{d}{2}}}
\nonumber \\
&&\frac{1}{(2 e B)^{\alpha-\frac{d}{2}}} \,\, \zeta(\alpha-\frac{d}{2},
\frac{M^2+e B}{2 e B}) \Bigg\vert_{\alpha=0,d=2}
\end{eqnarray}
\vskip 0.2 cm
where $\zeta(z,q)$ is the generalized Riemann zeta function
\begin{equation}
\label{four}
\zeta(z,q)=\sum_{n=0}^\infty \frac{1}{(q+n)^z}
=\frac{1}{\Gamma(z)} \int_{0}^{\infty} d t\,\,\frac{t^{z-1} e^{-q t}}{1-e^
-{t}}
\end{equation}
The potential (\ref{three}) has poles at $\alpha$=0, 1 and 2 for d=2
which can be absorbed in the counter terms. The finite part depends on
the exact renormalization conditions that are imposed. In what follows
we would use the $\overline{MS}$ renormalization scheme. From eqns. (\ref{three}) and 
(\ref{four}) we can write
\begin{equation}
\label{five}
V_{1}(M^2,B)= -\frac{e B}{32 \pi^2} \lim_{\alpha \rightarrow 0}
\frac{\partial}{\partial \alpha} \frac{(2 e B)^{1-\alpha}}{\Gamma(\alpha)}
\int dt\,\,t^{\alpha-2} \,\,\frac{{e}^{-\frac{M^2}{2 e B} t}}{\sinh 
\frac{t}{2}}
\end{equation}
which converges for Re $\alpha>2$. We analytically continue the result in the
complex $\alpha$-plane and use dimensional regularization technique to 
extract the finite contribution. To proceed further we first consider the 
case $\frac{M^2}{2 e B}<1$, expand $e^{-\frac{M^2}{2 e B} t}$ and formally 
integrate (\ref{five}) to obtain \cite{stegun} 

\begin{eqnarray}
\label{six}
V_{1}(M^2,B)&=&-\frac{\left\vert q \right\vert B}{32 \pi^2}\,\, 
\lim_{\alpha \rightarrow 0}\,\,
\frac{\partial}{\partial \alpha}\,\, \sum_{n=0}^\infty\,\, (\frac{M^2}
{2 e B})^{\alpha+\nu-1} \frac{(-1)^\nu}{\nu!\,\, (M^2)^{\alpha-1}} 
\nonumber \\
&&\frac{2}{\Gamma(\alpha)}(2^{\alpha+\nu-1}-1)\,\, \Gamma(\alpha+\nu-1) 
\,\,\zeta(\alpha+\nu-1)
\end{eqnarray}
Keeping leading terms in $\frac{M^2}{2 e B}$ we obtain
\begin{eqnarray}
V_{1}(M^2,B)&=&-\frac{1}{16 \pi^2}\,\,  [\,\,\frac{e^2 B^2}{2 \pi} \,\,  \zeta(2)  \log 2 e B +
\nonumber \\
&&\frac{e B M^{2}}{2}\,\, \log 2 - M^4 \frac{\pi}{2}\,\, \log
2 e B+..]
\end{eqnarray}
The leading term for the contribution of charged Goldstone bosons relevant
 for symmetry considerations is
\begin{equation}
V_{1}(M^2,B) \sim  -\frac{e B M^2}{32 \pi^{2}} \log 2
\end{equation}
which agrees with the earlier results \cite{linde2,smilga} upto a factor of order one.
For the case of $\frac{M^2}{2 e B}>1$ we write (\ref{five}) as
\begin{equation}
V_{1}(M^2,B)=-\frac{1}{32 \pi^2}\,\, \lim_{\alpha \rightarrow 0}\,\, \frac{\partial}
{\partial \alpha}\,\,  \frac{1}{\Gamma(\alpha)}\,\, \int_{0}^{\infty} {d x}\,\,  x^{\alpha-3}
e^{-M^{2} x} \frac{e B x}{\sinh e B x}
\end{equation}
and keeping leading terms obtain
\begin{equation}
V_{1}(M^{2},B) \simeq  \frac{1}{64 \pi^{2}} [M^{4} (\log M^{2}-\frac{3}{2})-\frac
{2}{3} (e B)^{2}\log M^{2}]
\end{equation}
which agrees with the result obtained by Salam and Strathdee \cite{salam2}.
 Likewise for the charged fermion fields using (\ref{one})  we obtain
\begin{equation}
V_{1}(M^{2},B)=\frac{4 \left\vert q \right\vert B}{32 \pi^{2}} \lim_
{\alpha \rightarrow 0} 
\,\, \frac{\partial}{\partial \alpha}\,\, \frac{(2 
\left\vert q \right\vert B)^{1-\alpha}}{\Gamma(\alpha)} \int_{0}^{\infty} d t\,\,  t^{\alpha-2}
 e^{-\frac{M^{2}}{2 \left\vert q \right\vert B} t}\coth \frac{t}{2} 
\end{equation}
The factor of 4 and positive sign account for the spinor nature of the fermi
field. In the limits mentioned above, we obtain
\begin{equation}
\label{ten}
V_{1}(M^{2},B)=\frac{\left\vert q \right\vert B M^{2}}{8 \pi^{2}}
 ( 1-\log M^{2})
\end{equation}
and
\begin{equation}
V_{1}(M^{2},B) \simeq -\frac{1}{16 \pi^{2}} [ M^4\,\, (\log M^2-\frac{3}{2})+\frac{2}{3}\,\, (\left\vert q \right\vert B)^{2} \log M^{2} ]
\end{equation}
for $\frac{2 \left\vert q \right\vert B}{M^2} > 1$ and $< 1$ respectively.

The total $ V_{eff}(\sigma,B)$ for the sigma model at the one loop
level is thus given  by

\begin{eqnarray}
\label{seven}
V_{eff}(\sigma,B) &=& -\frac{1}{2} \mu^2 \sigma^2+\frac{\lambda}{4} \sigma^4
\nonumber \\
&+&\frac{1}{64 \pi^2}\,\, (3 \lambda \sigma^2-\mu^2)^2\,\, \log\,\,
(\frac{3 \lambda\sigma^2-\mu^2}{m_\sigma^2}-\frac{3}{2}) \nonumber \\
&+& \frac{1}{64 \pi^2}\,\, (\lambda\sigma^2-\mu^2)^2\,\, \log\,\, (\frac{\lambda \sigma^2
-\mu^2}{m^2}-\frac{3}{2})
\nonumber \\
&-&\frac{e B}{16 \pi^{2}}\,\, (\lambda \sigma^2-\mu^2)\,\, \log 2
\nonumber \\
&-& \frac{N_{c}}{16 \pi^2} \sum_{flav}\,\, [g^{4} \sigma^{4}\,\,
(\,\, \log \frac{g^{2}\sigma^2}{m_{f}^2}-\frac{3}{2})
\nonumber \\
&+&\frac{2}{3}\,\, (\left\vert q \right\vert B)^{2}\,\, \log \frac{g^{2} \sigma^{2}}{
m_{f}^2}]
\end{eqnarray}

For $\frac{\left\vert q \right\vert B}{M^2} > 1$  ,the last term in eqn.(\ref{seven}) is replaced by eq. (\ref{ten}) summed
over flavors.
In figure 1, we plot $V_{eff}(\sigma,B)$ as a function of $\sigma$ for 
different values of magnetic field and compare it with the case of zero 
magnetic field by ignoring the B independent one loop terms. As input 
parameters we choose the constituent quark mass $m_f$= 500 MeV, sigma mass
$m_\sigma$=1.2 GeV and $f_{\pi}$=93 MeV.
We find that in the presence of intense magnetic fields the chiral symmetry breaking is enhanced. For magnetic field large compared to $m_f^2$ 
, from eqn.(\ref{ten})  we observe that though the fermionic contribution is towards symmetry
restoration, it is not enough to offset the contribution of charged goldstone pions.

              In order to study chiral symmetry restoration in the
 case of neutron stars as a 
function of chemical potential $\mu$ associated with finite baryon number density
we employ the imaginary time formalism by summing over Matsubara
frequencies.
This amounts to adding the fermionic free energy to the one loop
effective potential and is given by 
\begin{equation}
V_{1}^{\beta}(\sigma)=-\frac{\gamma}{\beta} \int \frac{d^3 k}{(2 \pi)^3}
\,\, \ln \,\,(1+e^{-\beta(E-\mu)})
\end{equation}
 which in the presence of static uniform magnetic field becomes 
\begin{equation}
V_{1}^{\beta}(\sigma)=-\frac{\gamma}{\beta}\,\, \frac{e B}{2 \pi} \sum_{n=0}^{\infty}
\int_{0}^{\infty} \frac{d k_{z}}{2 \pi}\,\, \ln\,\, (1+ e^{-\beta (E-\mu)})
\end{equation}
where $\gamma$ is the degeneracy factor and is equal to 2$N_c$ for each quark flavor.
We consider cold dense isospin symmetric quark matter for which the integrals can be
performed analytically.
The baryon number density corresponding to the chemical potential $\mu$ is given by
the usual thermodynamical relations.
\begin{equation}
N_{B}(\mu,0)=\frac{1}{3} \sum_{flav}\,\, \frac{\gamma}{6 \pi^{2}} \,\, (\mu^{2}-g^{2} \sigma^2)^{\frac{3}{2}}
\end{equation}
 and
\begin{equation}
N_{B}(\mu,B)= \frac{1}{3} \sum_{n=0}^{n_{max}}\,\,\frac{\gamma \left\vert
 q \right\vert B}{4 \pi^2}\,\, (2-\delta_{\mu,0})\,\, \sqrt{\mu^{2}-g^{2} 
\sigma^{2}-2 n \left\vert q \right\vert B}
\end{equation} 
for zero and finite magnetic field respectively.
Here $n_{max}$= Int $\left\vert \frac{\mu^2-g^2 \sigma^2}{2 \left\vert q
\right\vert B} \right\vert$.
To study chiral symmetry behavior at finite density in the presence of uniform
magnetic field, we minimize effective potential with respect to the
order 
parameter $\sigma$ for fixed values of chemical potential and magnetic field ( which then fixes the baryon density ). The results are shown in
figure 2 
where we have plotted the order parameter $\sigma$ as a function of
density 
at T=0 for different values of magnetic field. The solution indicates a first
order phase transition. The actual transition takes place at the point where
the two minima of the effective potential at $\sigma$=0 and 
$\sigma$=$\sigma$($\mu$,B) non zero become degenerate. The lower values of $\sigma$ (shown by dotted curves) are unphysical
in the sense that they do not correspond to the lowest state of energy. We find that magnetic field continues to enhance
chiral symmetry breaking at low densities as expected but as the magnetic field
is raised the chiral symmetry is restored at a much lower density compared to
the free field finite density case. This can be clearly seen from figure 3 where we have plotted the phase diagram in terms of baryon density and magnetic field.

         In conclusion we have examined the chiral symmetry behavior of
 the Linear Sigma model in the presence of static, uniform magnetic field
at the one loop level at zero density and at densities relevant in the
core of neutron stars. We find that the contribution of scalar and
fermion loops leads to an increase in chiral symmetry
breaking. At high densities too, this effect persists and for magnetic fields
of strength upto $10^{18}$ Gauss, there is enhancement in chiral symmetry 
breaking resulting in the restoration of symmetry at densities higher than if
no magnetic field were present. However, in the case of high
magnetic field $B\geq10^{19}$ Gauss the chiral symmetry is restored 
at lower densities.
              Thus in the core of neutron stars, if the
nuclear matter undergoes a transition to deconfined quark matter, the presence
of magnetic field would imply the existence of massive quark matter due to
enhancement in chiral symmetry breaking. This would affect the equation of 
state and will have astrophysical implications.

\begin{section}*{Acknowledgements}
We would like to thank Professor J.V. Narlikar for providing hospitality at
the Inter-University Centre for Astronomy and Astrophysics, Pune 411
007, India where this work was initiated. 
\end{section}

\pagebreak

Figure captions
\vskip 1 cm
Figure 1. Effective Potential in units of $(100 MeV)^4$ as a function of
$\sigma(MeV)$ for different values of the magnetic field. The curves a, b, c, d
and e are for B=0, $10^{16}, 10^{18}, 10^{19}$ and $3\times10^{19}$ Gauss
respectively.    

\vskip 0.5 cm
Figure 2a. Chiral condensate $\sigma(MeV)$ as a function of baryon density
in $f_m^{-3}$ for magnetic field B=0.

\vskip 0.5 cm
Figure 2b. Chiral condensate $\sigma(MeV)$ as a function of baryon density
in $f_m^{-3}$ for magnetic field B=$10^{16}$ Gauss.

\vskip 0.5 cm
Figure 2c. Chiral condensate $\sigma(MeV)$ as a function of baryon density
in $f_m^{-3}$ for magnetic field B=$10^{18}$ Gauss.

\vskip 0.5 cm
Figure 2d. Chiral condensate $\sigma(MeV)$ as a function of baryon density
in $f_m^{-3}$ for magnetic field B=$10^{19}$ Gauss.

\vskip 0.5 cm
Figure 3. Phase diagram as a plot of magnetic field  versus baryon density
in units of $fm^{-3}$.
\pagebreak








\begin{thebibliography}{99}
\bibitem{brown}  See for example, A.~V.~Smilga, Phys. Rep. $\mathbf{291}$,
                 1 (1998);  
                 G.~E.~Brown and M.~Rho, Phys. Rep.
                $\mathbf{269}$,
                333 (1996) and references cited therein.
\bibitem{harris} J.~W.~Harris and B.~M\"{u}ller,Ann. Rev. Nucl. Part. Sci.
                $\mathbf{46}$,71 (1996).
\bibitem{linde1} See for example, A.~D.~Linde, Rep. Prog. Phys. $\mathbf{42}$
                 , 289 (1979).
\bibitem{salam1}  A.~Salam and J.~Strathdee, Nature $\mathbf{252}$,
                  569 (1974).
\bibitem{salam2}  A.~Salam and J.~Strathdee, Nucl. Phys.
                 $\mathbf{B90}$, 203 (1975).
\bibitem{linde2}  A.~D.~Linde, Phys. Lett. B $\mathbf{62}$, 435 (1976).
\bibitem{kikuchi} Y.~J.~Ng and Y.~Kikuchi, in {\it Vacuum  Structure  in
                  Intense  Fields}, edited by H.~M.~Fred and B.~M\"{u}ller
(Plenum, NewYork, 1991); D.~M.~Gitman, E.~S.~Fradkin
and Sh.~M~Shvartsman, in {\it Quantum  electrodynamics
with  unstable  Vacuum}, edited by V.~L.~Ginzburg(Nova Science,
Commack,NY. 1995).  
\bibitem{ternov} I.~M.~Ternov and O.~F.~Dorofeev, Phys. Part.Nucl. $\mathbf{25}$,
                 1 (1994); J.~Daicic, N.~E.~Frankel and V.~Kovalenko, Phys. Rev.
Letts., $\mathbf{71}$, 1779 (1993); C.~Thompson and R.~C.~Duncan, Astrophys. Jour.,
                  $\mathbf{408}$, 194 (1993); ibid $\mathbf{473}$, 322 (1996);
                     C.~Kouveliotou et.al, Nature  $\mathbf{393}$, 235 (1998).

\bibitem{bocquet} M.~Bocquet, S.~Bonazzola, E.~Gourgoulhon and J.~Novatz,
                  Astrophys, $\mathbf{301}$, 759 (1995) ; C.~Thompson
                   and B.~C.~Duncan, astrophys, J.  $\mathbf{392}$,
                    19 (1992) ; J.~Daicic, N.~E.~Framkel and
                  V.~Kovalinko, Phys. Rev. Lett. $\mathbf{71}$,
                  1979 (1995).
\bibitem{miransky} V.~A.~Miransky,{\it Dynamical  Symmetry  Breaking  in
Quantum  Field  Theory} (World Scientific, Singapore, 1993).
\bibitem{klevansky} S.~P.~Klevansky and R.~H.~Lemmer Phys. Rev. D$\mathbf{39}$,
                   3478 (1989).

\bibitem{klimenko} K.~G.~Klimenko, Z Phys. C$\mathbf{54}$, 323 (1992);
                   K.~G.~Klimenko, A.~S.~Vshivtsev and B.~V.~Magnitsky, Nuovo
                   Cim. A$\mathbf{107}$, 439 (1994); S.~Schramm, B.~Muller and A.~J.~Schramm,
                   Mod. Phys. Lett. A$\mathbf{7}$, 973 (1992).
  
\bibitem{smilga}   I.~A.~Shushpanov and A.~V.~Smilga, Phys. Lett.B
                  $\mathbf{16}$ 402, 351 (1997).

\bibitem{gellmann} M.~Gell-Mann and M.~Levy, Nuovo Cimento $\mathbf{16}$,
                 705 (1960).
\bibitem{kirshnitz} D.~A.~Kirshnitz and A.~D.~Linde, phys. Letts. 
                    B$\mathbf{42}$, 471
                    (1972); S.~Wienberg, Phys. Rev. D$\mathbf{9}$,
                    3357 (1974);
                    L.~Dolan and R.~Jakiew, Phys. Rev. D$\mathbf{9}$,
                    3320 (1974);
                    C.~Bernard, Phys. Rev. D$\mathbf{9}$, 3312 (1974).
\bibitem{wu-yang}   See for example, Wu-Yang Tsai, Phys. Rev. 
                    D$\mathbf{7}$, 1945 (1973).
\bibitem{stegun}   {\it Handbook  of  Mathematical  functions}, edited
by M.~Abramowitz and I.~A.~Stegun (Academic Press, NewYork, 1965 ).
\end{thebibliography}
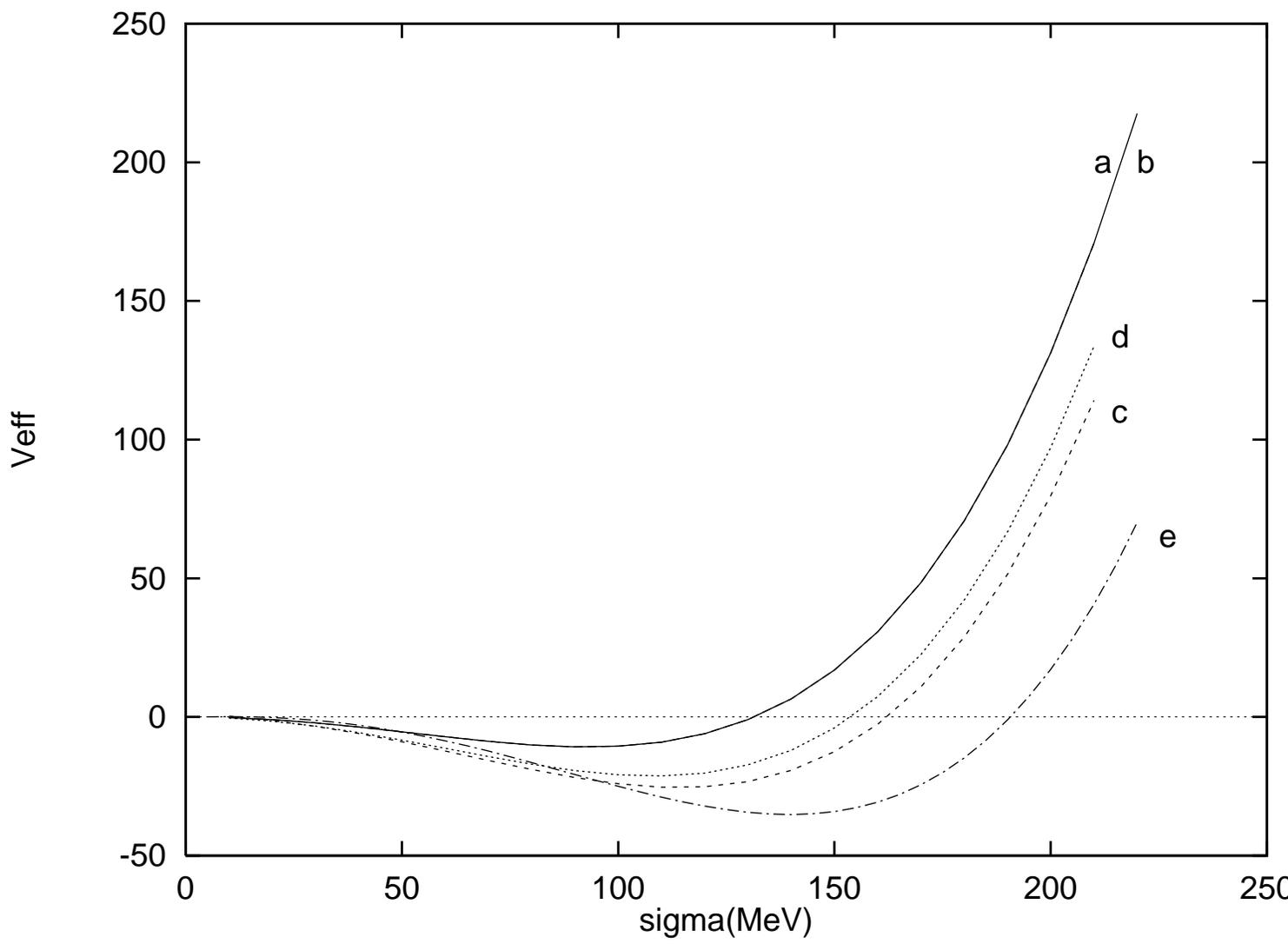
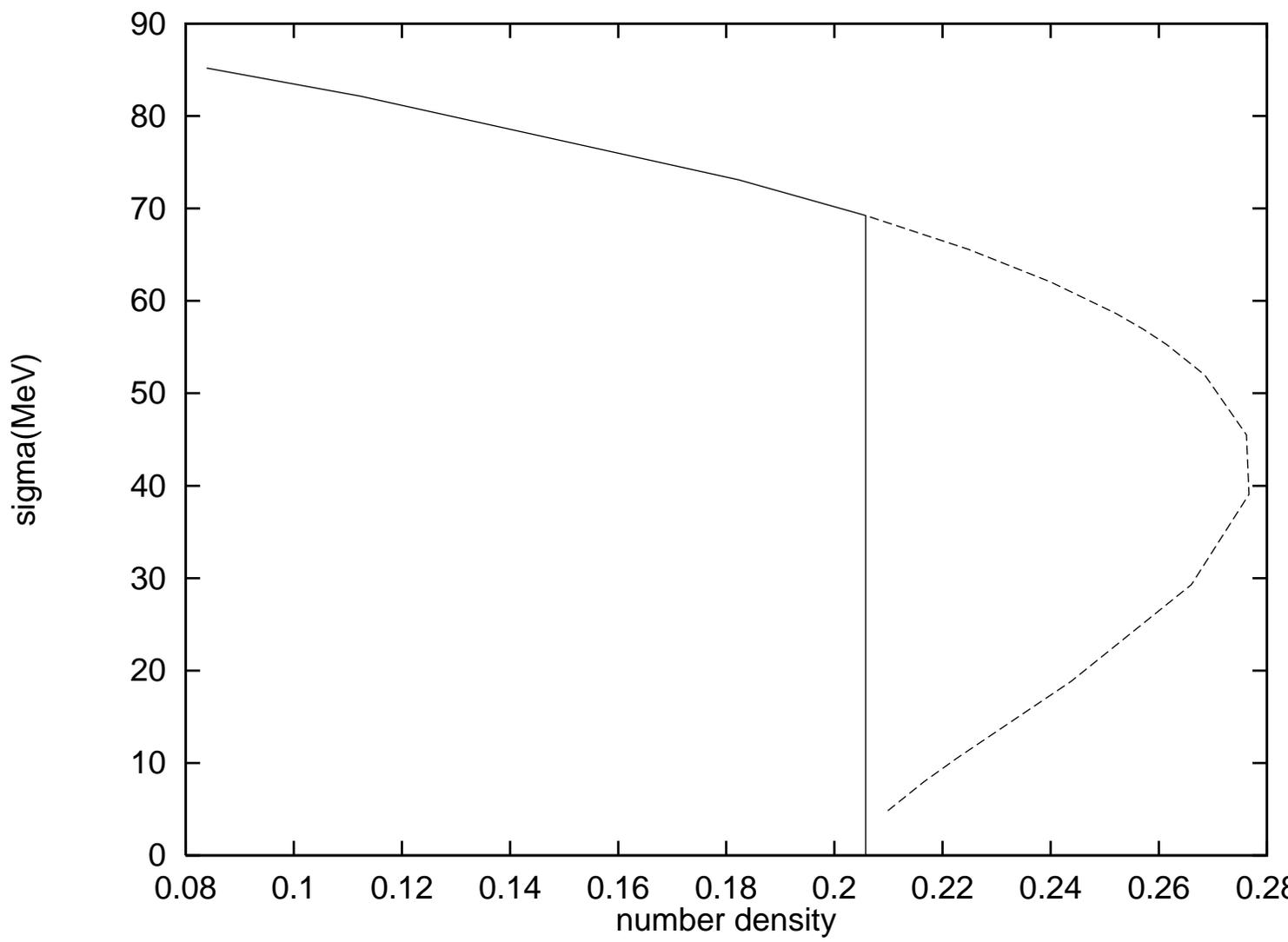
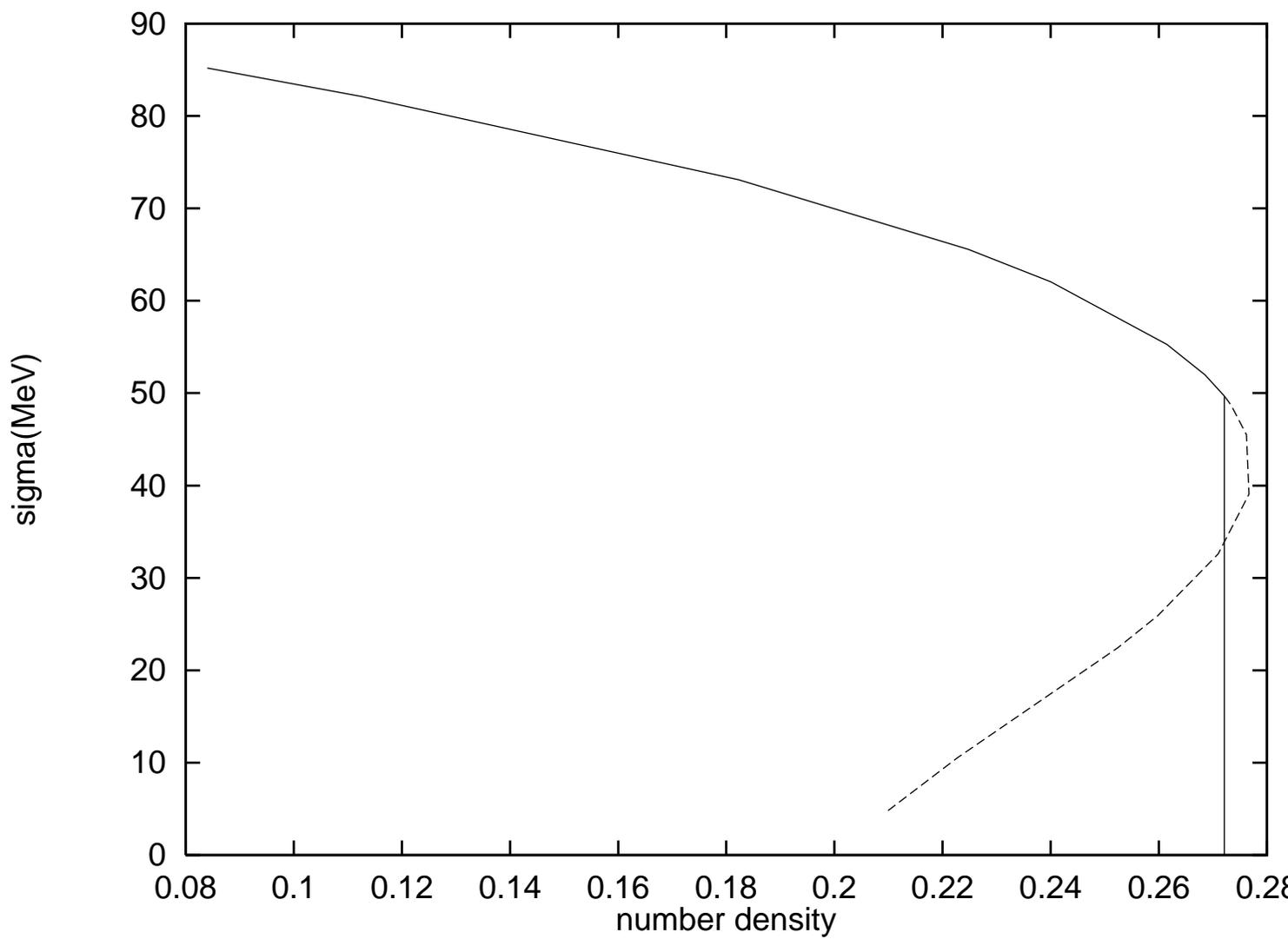
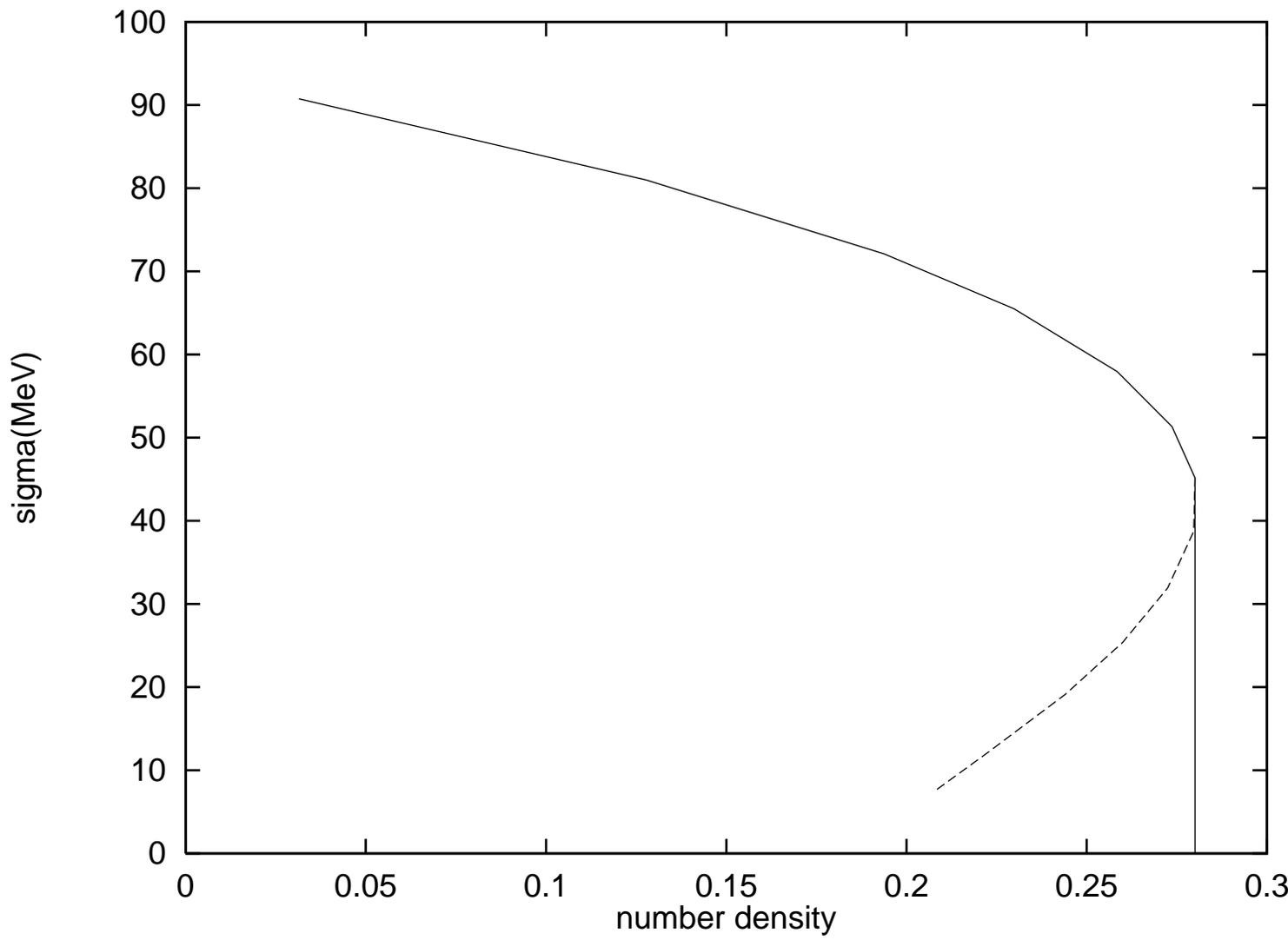
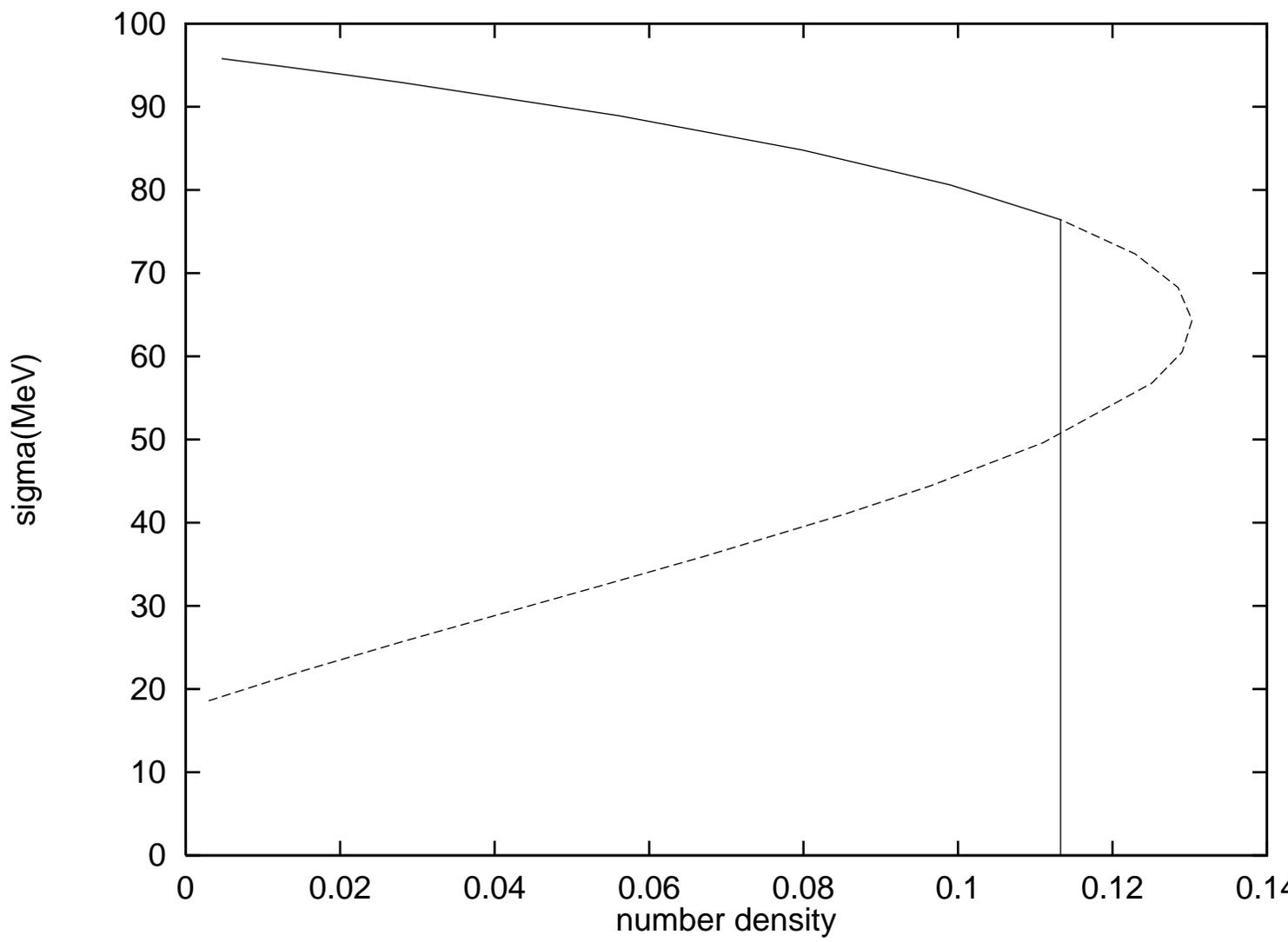
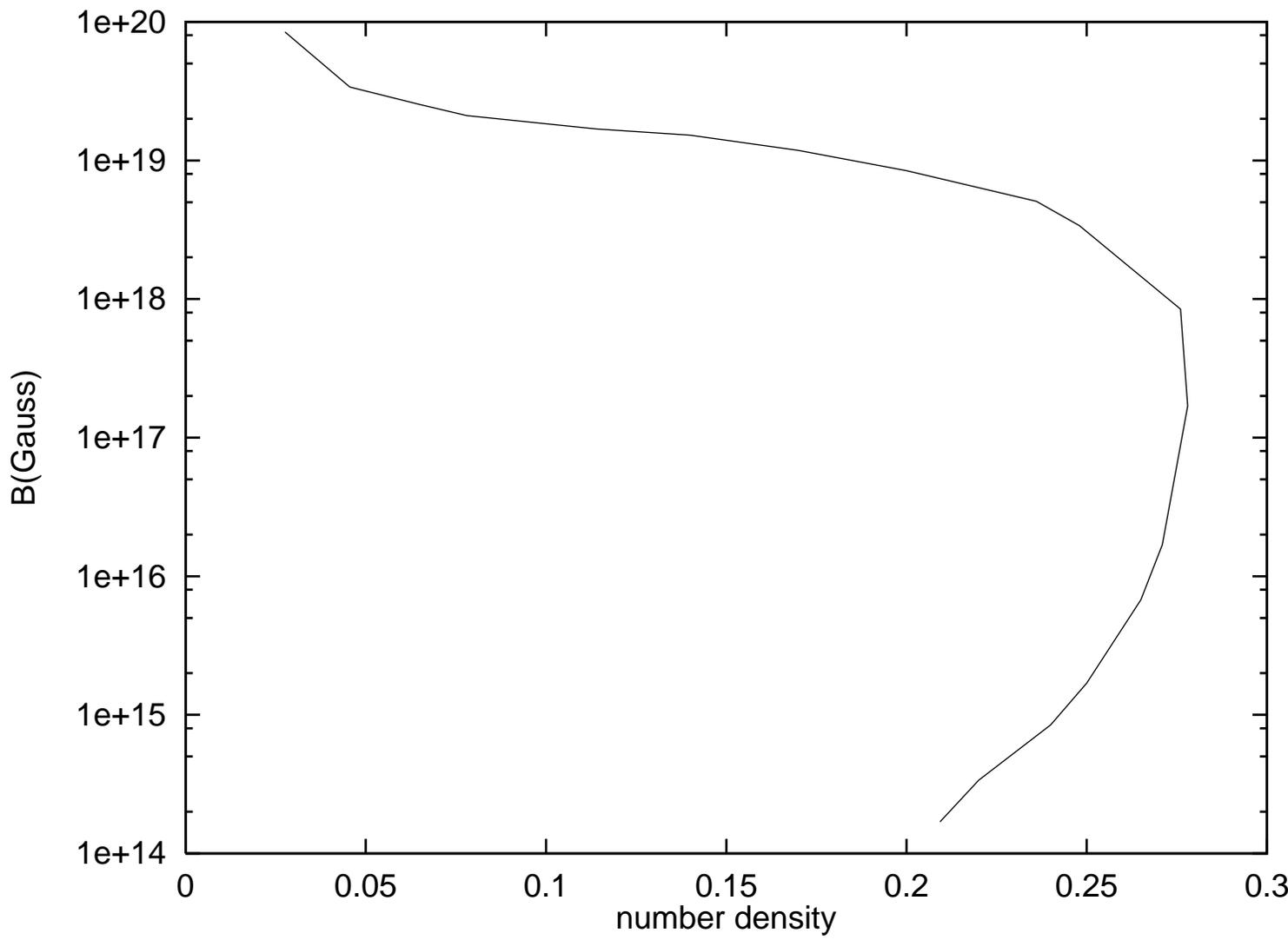
\end{document}